\documentclass[
reprint,
superscriptaddress,
showpacs,
nofootinbib,
amsmath,amssymb,
aps,
pra,
floatfix,longbibliography
]{revtex4-2}
\usepackage{graphicx}
\usepackage{dcolumn}
\usepackage{float}
\usepackage{booktabs}
\usepackage[hypertexnames, breaklinks=true, bookmarksnumbered=true, bookmarksopen=true, colorlinks=true, linktocpage=true, citecolor=blue, urlcolor=magenta, linkcolor=magenta]{hyperref}
\usepackage{amsmath,amssymb,amsfonts,amsthm}
\usepackage{mathtools}
\usepackage[T1]{fontenc}
\usepackage[latin9]{inputenc}
\usepackage{txfonts}
\usepackage{bbm}
\usepackage{bm}
\usepackage{mathrsfs}
\usepackage{xcolor}
\usepackage{natbib}
\usepackage{siunitx}

\usepackage{appendix}
\usepackage{makecell}
\begin{document}

\title{Bridging the classical and quantum regimes in a dissipative Ising chain}

 \author{Zhenming Zhang}
  \affiliation{Laboratory of Quantum Information, University of Science and Technology of China, Hefei 230026, China}
 \author{Haowei Li}
  \affiliation{Laboratory of Quantum Information, University of Science and Technology of China, Hefei 230026, China}
\author{Wei Yi}
\email{wyiz@ustc.edu.cn}
 \affiliation{Laboratory of Quantum Information, University of Science and Technology of China, Hefei 230026, China}
\affiliation{Anhui Province Key Laboratory of Quantum Network, University of Science and Technology of China, Hefei 230026, China}
\affiliation{CAS Center For Excellence in Quantum Information and Quantum Physics, Hefei 230026, China}
\affiliation{Hefei National Laboratory, University of Science and Technology of China, Hefei 230088, China}
\affiliation{Anhui Center for Fundamental Sciences in Theoretical Physics, University of Science and Technology of China, Hefei 230026, China}

\begin{abstract}
We study the long-time dynamics of a dissipative Ising chain with varying quantum correlation.
Invoking an ensemble-average formalism, and assuming spatial translation symmetry, we show that the dynamics can be described by a Lindblad master equation with an interpolated coherent Hamiltonian.
In the classical limit, the interpolation Hamiltonian leads to a set of nonlinear equations of motion, where limit cycles can emerge in the long-time dynamics.
In the quantum limit, by contrast, the system approaches a ferromagnetic steady state at long times.
In between the two extremes, the discrete spatial translation symmetry can be spontaneously broken, as an antiferromagnetic steady state emerges, bridging the classical and quantum regimes.
In particular, we illustrate how the classical limit-cycle behavior gradually disappears with the increase of quantum correlation.
Since our model in the two extremes respectively applies to a dissipative Rydberg gas in the high- and zero-temperature limits,
we expect it to provide a qualitatively correct description of dissipative Rydberg gases at interim temperatures, and shed light on the fate of limit cycles in a quantum open system.
\end{abstract}

\maketitle

\section{Introduction}
Dissipative Ising chains are fundamentally important for the study of many-body quantum open systems.
On one hand, the extensive study of non-dissipative Ising chains~\cite{Mbeng2024, SachdevQPT} lends valuable insights to their dissipative counterparts, wherein the form of dissipation can often be engineered for state preparation or exotic dynamics~\cite{Zhang2025, Morigi2015, Ates2012, Jin2018, Shibata2019}.
On the other hand, these models naturally arise in many state-of-the-art experimental systems such as trapped ions~\cite{Barreiro2011, Schneider2012}, cold atoms~\cite{Orth2008, Schwager2013}, and Rydberg gases~\cite{Saffman2010, ZDS2021, Browaeys2020, Labuhn2016, Zeiher2017}.
Since these systems are promising candidates for quantum simulation and computation, understanding the dynamics of dissipative Ising chains also has potential applications.

Dynamics of the dissipative Ising chain are well-studied in two limits. In the classical limit, one can adopt a mean-field description for the interactions.
This gives rise to nonlinear equations of motion, which further lead to remarkable nonlinear phenomena, such as bistability~\cite{Sibalic2016, deMelo2016} and limit cycles~\cite{ Cabot2024, Chan2015, Landa2020, Lee2011}.
Physically, the relevant phenomena have stimulated much interest in Rydberg vapors~\cite{ Urvoy2015, Carr2013, Ding2020, Wadenpfuhl2023, Ding2024, Wu2024, WuKD2024, Liu2024a, Liu2025b}, and have been investigated in the diverse contexts of
dissipative time crystals~\cite{Wilczek2012, Sacha2018, Ding2024, Wu2024, Liu2024a, Liu2025b},
dynamic phase transitions~\cite{Lee2011, Ates2012, Ding2020, Marcuzzi2014},
stochastic resonances~\cite{Gammaitoni1998, WuKD2024} and sensing~\cite{Degen2017, Yuan2023},
and non-Hermitian physics~\cite{Ashida2020, Lourenco2022}.
By contrast, in the quantum limit, the system is dominated by quantum correlations, with intriguing transient phenomena such as collective jumps~\cite{Lee2012} and quantum stochastic resonances~\cite{ Li2024}, in the steady-state approaching dynamics governed by the quantum master equation.
These phenomena are expected to manifest in dissipative Rydberg gases or Rydberg-atom arrays at low enough temperatures.
In an effort to understand the two limits in a unified framework, previous studies often adopt cluster models~\cite{Yamamoto2009, Jin2016, Jin2018, Huybrechts2019}, where the intra-cluster interactions remain intact and quantum mechanical, while the inter-cluster interactions are treated using the mean-field approximation.
However, while the cluster model can recover the classical non-linear descriptions by setting the cluster size to one atom, a slight increase in the cluster size would drastically modify the mean-field behaviour, especially the limit cycle dynamics~\cite{Jin2018,LiJin2021,LiJin2023}.
Further, the cluster models immanently reduce the translational symmetry that persists in both the classical and quantum limits. Hence, while the model is capable of capturing the beyond-mean-field effects, its ability to connect the classical and quantum regimes is limited.

Aiming to bridge the two limits, in this work, we study the long-time dynamics of dissipative Ising chains through an interpolated model. The model is derived as an ensemble-averaged cluster model, which manifestly interpolates between the non-linear equations of motion and the quantum master equation, without breaking the spatial translation symmetry.
In the classical limit, our model recovers the limit-cycle behavior, while in the quantum limit, the system approaches a ferromagnetic steady state at long times~\cite{Weimer2015, Honing2013}.
More importantly, we show that the system features an antiferromagnetic steady state in between the two limits. Starting from the classical limit, with the increase of quantum correlation, Hopf bifurcations~\cite{Strogatz1994} occur, as oscillatory dynamics of the limit cycle give way to to antiferromagnetic steady state at long times. The steady state eventually becomes ferromagnetic through a steady-state phase transition as the system is tuned toward the quantum limit.
Our results thus shed light on the fate of limit cycles in the presence of quantum correlation, and
are relevant to the on-going study of dissipative Rydberg gases.

This work is organized as follows. In Sec.~II, we derive the interpolated model by performing an ensemble average of cluster models. In Secs.~III and IV, we map out the phase diagram and characterize the steady-state phase transitions, respectively. In Sec.~V, rich nonlinear behaviors are discussed, including limit-cycle dynamics and frequency bifurcations.
Finally, we summarize in Sec.~VI.

\section{Model}
We consider a dissipative one-dimensional Ising chain described by the Lindblad master equation
\begin{equation}\label{masterEQ}
    \dot \rho = -i[H,\rho]+\gamma\sum_j\left(L_j\rho L_j^\dagger -\frac{1}{2}\{L_j^\dagger L_j,\rho\}\right),
\end{equation}
where the coherent Hamiltonian is given by
\begin{align}
    H = &\sum_j  \left(-\Delta n_j  +\frac{\Omega}{2} \sigma^x_j\right) +H_{\text{int}},
\end{align}
with the quantum jump operators $L_j = \sigma^-_j$.
Here the subscript \( j \) labels the individual spins in the chain, $\sigma_j^{x,y,z}$ are the Pauli operators of the $j$th spin, the number operator \( n_j = (\sigma_j^z + 1)/2 \),
and $\Delta$, $\Omega$, and $\gamma$ are parameters.
The interaction of the Ising chain is given by
\begin{equation}
H_{\text{int}} = V\sum_{i}  n_i n_{i+1},
\end{equation}
where $V$ describes the nearest-neighbor interaction strength.

Physically, the dissipative Ising chain above can be realized in a dissipative Rydberg-atom array~\cite{Saffman2010, ZDS2021, Browaeys2020, Labuhn2016, Hoening2014, Morigi2015, Letscher2017}, where the ground and Rydberg state of the $j$th atom form the basis of the $j$th spin. The quantum-jump processes in Eq.~(\ref{masterEQ}) correspond to the spontaneous decay of the Rydberg states, which satisfy the Born-Markov approximation~\cite{opentextbook}, justifying the application of the Lindblad master equation. Indeed, such a treatment is widely used for describing dissipative Rydberg atoms~\cite{ Urvoy2015, Carr2013, Ding2020, Wadenpfuhl2023, Ding2024, Wu2024, WuKD2024, Liu2024a, Liu2025b}.
It also follows that, $\Omega$ is the effective Rabi frequency of the coupling between the ground and Rydberg states, $\Delta$ is the detuning of the coupling, and $n_j$ corresponds to the number operator for the Rydberg state of the $j$th atom.
The interaction derives from the Van der Waals interaction between adjacent Rydberg atoms.

The model above and the master equation (\ref{masterEQ}) have been extensively studies, mostly in two limits.
In the quantum limit, a regime associated with Rydberg arrays at low enough temperatures~\cite{ZDS2021, Hoening2014, Morigi2015, Letscher2017},
the Lindblad equation (\ref{masterEQ}) is treated exactly, under which the system exhibits collective dynamics and asymptotically approaches a steady state at long times.
In the classical limit, a regime associated with Rydberg vapors above the room temperature, a mean-field approximation is usually adopted, leading to nonlinear equations of motion.
The resulting rich nonlinear dynamics, such as the bistability transition, limit cycles (often dubbed time crystals in recent years), and Hopf bifurcations, have been reported in many recent experiments~\cite{Carr2013, Wu2024, Ding2024, Wadenpfuhl2023}.

To describe the interim regime, previous studies have introduced the cluster model, where various fully quantum mechanical clusters exist along the chain, while the inter-cluster interactions are treated on the mean-field level~\cite{Yamamoto2009, Jin2016, Jin2018, Huybrechts2019}. When the cluster size is uniformly one spin (or one atom),
the model reduces to that under the mean-field approximation, leading to nonlinear dynamics. However, with the increase of cluster size, the nonlinear behaviors can become drastically modified. Further,
by dividing the system into finer clusters, the cluster model reduces the translational symmetry of the system in both the classical and the quantum limits.
These observations suggest that the cluster model has limited capability in connecting the two limits.

For this purpose, we consider an ensemble of systems with different cluster partitions.
For a one-dimensional system of \( N \) spins with nearest-neighbor interactions under the periodic boundary condition, 
there are a total of $N$ local interaction terms $h_{i,i+1}$, and each can be switched on or off, leading to \( 2^N \) possible partitions. Each partition can be uniquely labeled by a set \( C_j = \{ j_1, j_2, \ldots \} \), where the element \( j_i \) represents the bond on which the local interaction \( h_{j_i,j_i+1} \) is replaced by its mean-field decoupled counterpart \( h_{j_i,j_i+1}^{\text{MF}} \). Here
\begin{align}
  h_{i,i+1}&=V  n_i  n_{i+1},\\
   h_{i,i+1}^{\text{MF}} &= V\Big( \langle n_i \rangle n_{i+1} + n_i \langle n_{i+1} \rangle - \langle n_i \rangle \langle n_{i+1} \rangle \Big).
\end{align}

The Hamiltonian of a given partition \( C_j \) is given by
\begin{equation}
    H(C_j) = H_0 +\sum_{i} \xi_{i,j} h_{i,i+1}+ \sum_{i}(1-\xi_{i,j}) h_{i,i+1}^{\text{MF}},
\end{equation}
where \( H_0 =\sum_j \left(-\Delta n_j +\frac{\Omega}{2} \sigma^x_j\right)\) is the partition-independent, non-interacting Hamiltonian. Following the spirit of cluster models, we further require \( \xi_{i,j}=0 \) for \( i \in C_j \), and \( \xi_{i,j}=1\)  for \( i \notin C_j \).

To proceed, we make the following assumptions:
\begin{enumerate}
    \item \textbf{Ensemble average over all partitions.}
    We consider an ensemble of systems, where each evolves under a given partition. The probability distribution of partition $C_j$ in the ensemble is given by $P(C_j)$. We assume that the evolution of the density matrix is given by the ensemble average of the evolutions within different partitions (that is, governed by different realizations of the cluster model). Note that, in the quantum limit, \( P(\emptyset) = 1 \), and in the classical limit, \( P(\{1,2,\dots,N\}) = 1 \), which recover the equations of motion in either case.
    \item \textbf{Translational symmetry.}
    We assume the probability distribution respects the translational symmetry: \( P(C_j) = P(C'_j)\), where \( C'_j \) is a set obtained by shifting each element in \( C_j \) by one.
    This assumption is inspired by the fact that the translational symmetry is present in both the classical and quantum limits.
\end{enumerate}

With these, the Lindblad master equation corresponding to partition $C_j$ is given by
\begin{equation}
\dot{\rho} = -i[H(C_j), \rho] + \gamma \sum_i \left( L_i \rho L_i^\dagger - \frac{1}{2} \{ L_i^\dagger L_i, \rho \} \right) := \mathcal{L}_j \rho,
\end{equation}
where $\mathcal{L}_j$ denotes the Liouvillian superoperator associated with $C_j$.
According to the ensemble average assumption, the averaged density matrix evolves according to
\begin{align}
\dot \rho
= & \sum_{j} P(C_j) \mathcal{L}_{j} \rho \nonumber \\
= & -i \sum_{j}P(C_j) [H(C_{j}), \rho] + \gamma  \sum_i \left( L_i \rho L_i^\dagger - \frac{1}{2} \{ L_i^\dagger L_i, \rho \} \right).
\end{align}
It follows that the system effectively evolves under the effective coherent Hamiltonian
\begin{align}
H_{\text{eff}}  &= \sum_j P(C_j) H(C_j)\\
    &=H_0  + \sum_{i,j}P(C_j)\xi_{i,j} h_{i,i+1} + \sum_{i,j}P(C_j)(1-\xi_{i,j}) h_{i,i+1}^{\text{MF}}.
\end{align}
Given the translational invariance of the probability distribution \( P(C_j) \), that is, \( \sum_j P(C_j) \xi_{i,j} = \sum_j P(C_j) \xi_{i+1,j} \), the summation $\sum_j  P(C_j) \xi_{i,j}$ is independent of $i$, which we denote as \( \lambda \).
We thus derive an effective Hamiltonian that interpolates between the two limits
\begin{equation}
H_{\text{eff}} = H_0 + \lambda \sum_i h_{i,i+1} + (1 - \lambda) \sum_i h_{i,i+1}^{\text{MF}}.\label{eq:interpolation}
\end{equation}
Note that for the derivation above, the periodic boundary condition is imposed.
As the system is tuned from the classical limit toward the quantum limit, \( \lambda \) increases smoothly from $0$ to $1$. Hence $\lambda$ can be regarded as a measure of quantum correlation.
In particular, in the limit $\lambda=0$, the model is equivalent to substituting the product state $\rho = \bigotimes_j \rho_j$ into Eq.~(\ref{masterEQ}), under which there are no entanglements between spins on different sites~\cite{Li2024}.
Note that we have not systematically introduced temperature nor considered the thermal effects. But, since our model in the two limits $\lambda=0$ and $\lambda=1$ respectively recovers the widely used descriptions for dissipative Rydberg atoms at high and zero temperatures~\cite{Lee2011,Carr2013,Weimer2010,WLee2019}, and since quantum correlations become increasingly suppressed at higher temperatures, we expect our calculations in the intermediate regime ($0<\lambda<1$) should provide a qualitatively correct description for dissipative Rydberg gases at finite temperatures. A full-fledged finite-temperature theory is desirable but beyond the scope of current work.

\section{Dynamic phase diagram}
We now study the long-time dynamics of the system using the Lindblad master equation with the interpolation Hamiltonian (\ref{eq:interpolation}). Throughout the work, we set \( \gamma=1 \) as the unit of energy.
We focus on the parameters where the system exhibits limit cycles in the classical limit.
For a typical illustration, we take \( \Omega= 1.5 \), \( \Delta = 2 \), and \( V=5 \).

\begin{figure}[tbp]
\centering
\includegraphics[width=\columnwidth]{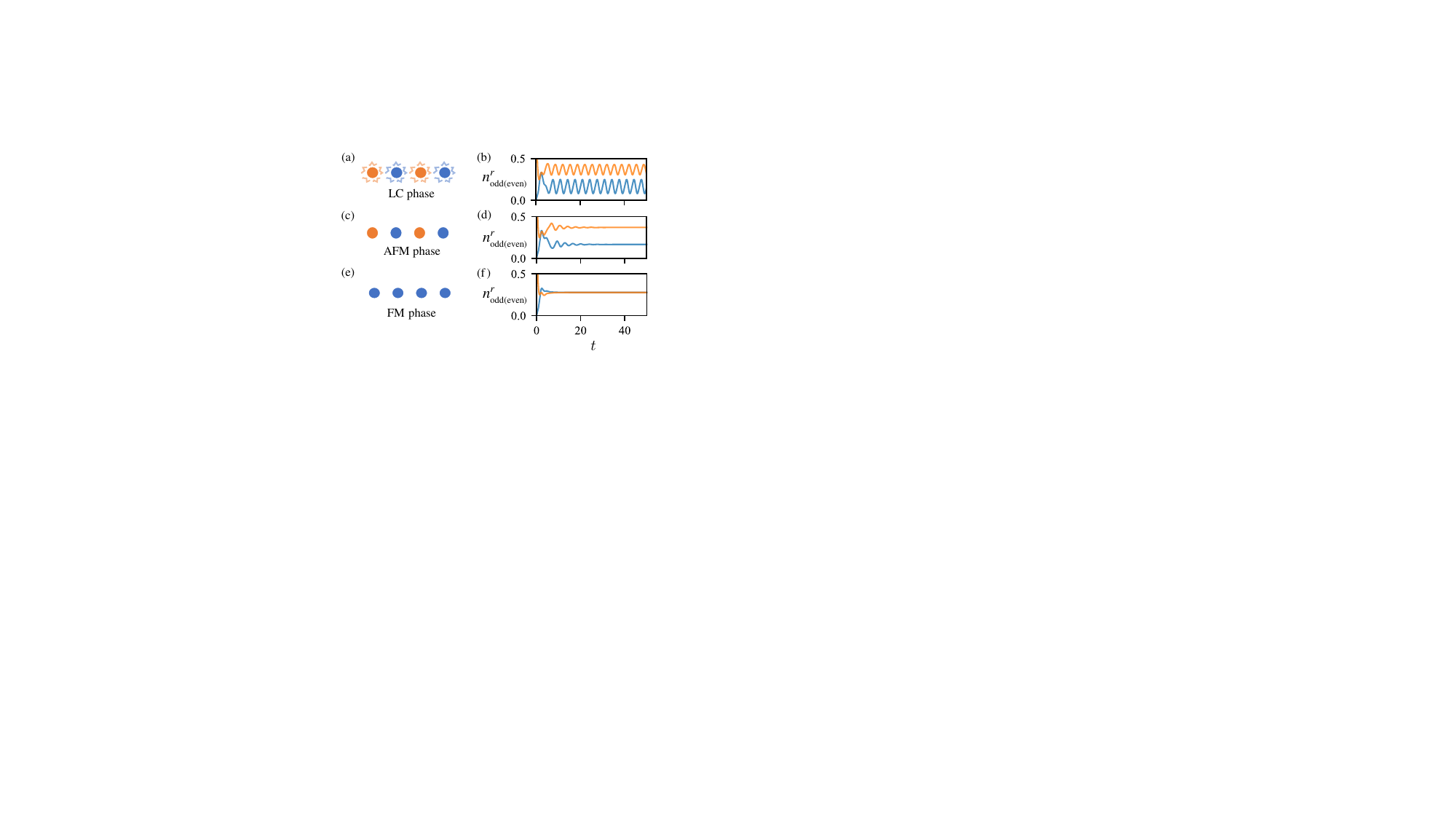}
\caption{ Schematic illustration and typical dynamics of the dynamical phases. (a)(c)(e)
Schematic illustration of the limit cycle phase (LC), the antiferromagnetic phase (AFM), and the ferromagnetic phase (FM), respectively.
Colors of the symbols indicate different spin-up occupations at long times.
(b)(d)(f) Numerically simulated time evolution of \( n_{\text{odd}}^r \) (orange line) and \( n_{\text{even}}^r \) (blue line) under the parameters \( \lambda = 0.1, 0.2, \) and \( 0.5 \), which correspond to the LC, AFM, and FM phases, respectively.
Other parameters are  \( \Omega=1.5 \), \( \Delta = 2 \), \( V=5 \), and \( N=6 \). }
\label{fig:example}
\end{figure}

By varying the parameter \( \lambda \), we identify three distinct dynamical phases: the limit cycle phase, schematically illustrated in Fig.~\ref{fig:example}(a) and denoted as LC; the antiferromagnetic phase, denoted as AFM as in Fig.~\ref{fig:example}(c); and the ferromagnetic phase, denoted as FM in Fig.~\ref{fig:example}(e).
In the LC phase, the system oscillates persistently rather than converging to a steady state.
The AFM phase is characterized by a steady state where the system spontaneously breaks the translational symmetry, with the spin-up population (or the Rydberg-state population) alternating between adjacent sites along the chain.
The FM state is characterized by a steady state with uniform spin-up (Rydberg-state) population.

For numerical simulations, we define the spin-up occupation on the $j$th site as \( n_j^r = \langle n_j \rangle \), where the expectation value is taken with respect to the density matrix \( \rho(t) \).
We then calculate the average spin-up occupation for odd lattice sites, $n^r_{\text{odd}}=\frac{2}{N}\sum_{i=1}^{N/2}n^r_{2i-1}$, and even lattice sites, $n^r_{\text{even}}=\frac{2}{N}\sum_{i=1}^{N/2}n^r_{2i}$, respectively.
Figure~\ref{fig:example}(b), (d), and (f) show the time evolution of \( n_{\text{odd}}^r \) and \( n_{\text{even}}^r \) for \( \lambda = 0.1, 0.2, \) and \( 0.5 \), respectively.
These numerical results are obtained for a chain of \( N=6 \) atoms with a periodic boundary condition.
For \( \lambda \lesssim 0.17 \), the system is in the LC phase, marked by sustained oscillations of \( n_j^r \) over time, typical of a continuous time crystal. When \( \lambda \) is in the range \( 0.17 \leq \lambda \leq 0.43 \), the system is in the AFM phase. For \( \lambda \gtrsim 0.43 \), the system enters the FM phase, where all atoms reach the same steady-state value of \( n_j^r \).

\begin{figure}[tbp]
 \centering
\includegraphics[width=\columnwidth]{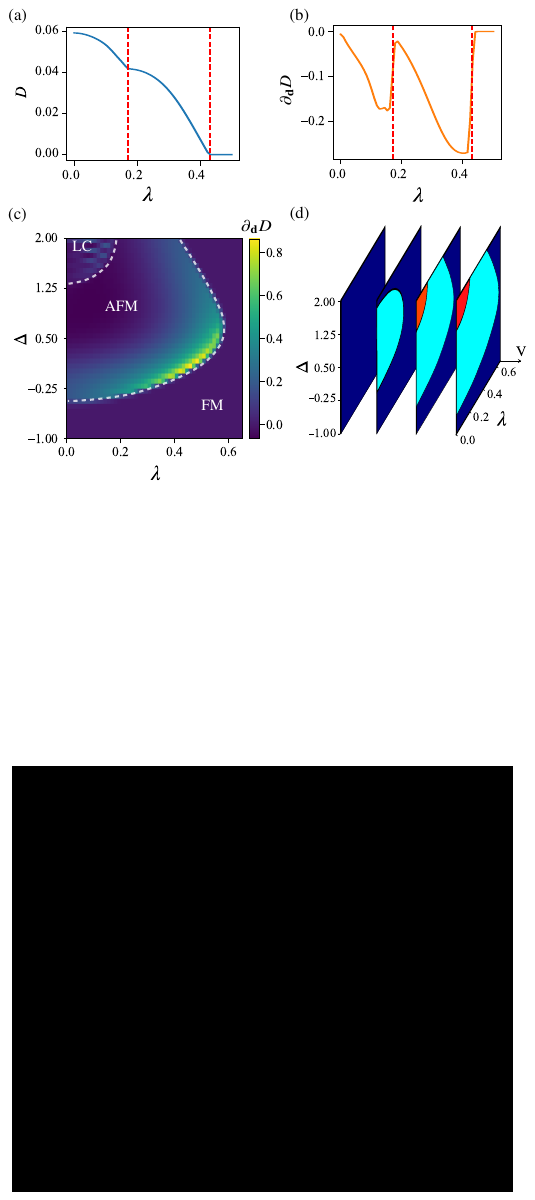}
 \caption{%
(a)(b) Numerically evaluated \( D \) and its directional derivative \( \partial_{\mathbf{d}} D \) along the direction \( \mathbf{d} = (1, 0) \), 
which is relatively close to the normal directions of both phase boundaries at $\Delta=2$. 
The red vertical dashed lines indicate the transition points between different dynamic phases. For the calculations here, we fix \( \Delta = 2 \) and \( V = 5 \).
(c) The directional derivative \( \partial_{\mathbf{d}} D \) with \( \mathbf{d} = \frac{1}{\sqrt{2}}(-1, 1) \), plotted as a function of both \( \lambda \) and \( \Delta \) for \( V = 5 \). Here $\mathbf{d}$ is chosen to align with the dominant normal direction of both phase boundaries. 
The white dashed lines denote the phase boundaries, determined by discontinuities in $\partial_{\mathbf{d}} D$, as illustrated (b).
(d) Phase diagrams in the \( (\lambda, \Delta) \) plane for different values of \( V \), where the phase boundaries are again identified by discontinuities in $\partial_{\mathbf{d}} D$. The dark blue region represents the FM phase, the cyan region corresponds to the AFM phase, and the red region indicates the LC phase. From left to right, the panels show the results for \( V = 1.5, 3, 4.5, \) and \( 6 \), respectively. Other parameters are fixed at \( \Omega = 1.5 \) and \( N = 6 \).}
\label{fig:phase diagrams}
\end{figure}

To further characterize the dynamic phase diagram, we define the variance
\begin{equation}
    D = \mathrm{Var}\left(\{n_j^r(t)  | j=1,2,..N, t\in\{t_i\}\} \right),
\end{equation}
where $\{t_i\}$ are equal-spaced discrete times after a sufficiently long time evolution. Here the interval between adjacent $t_i$ should be much smaller than the evolution time.

In Fig.~\ref{fig:phase diagrams}(a)(b), we show the function \( D \) and its directional derivative \( \partial_{\mathbf{d}} D \) along the direction \( \mathbf{d} = (1, 0) \), corresponding to increasing \( \lambda \). The vector \( \mathbf{d} \) is defined in the two-dimensional parameter space \( (\lambda, \Delta) \), where the horizontal axis represents \( \lambda \) and the vertical axis represents \( \Delta \). While \( D \) varies continuously across the phase boundaries, its directional derivative \( \partial_{\mathbf{d}} D \) exhibits clear discontinuities at the boundaries (highlighted by red dashed lines).
This is more visible in Fig.~\ref{fig:phase diagrams}(c), where we compute the directional derivative along \( \mathbf{d} = \frac{1}{\sqrt{2}}(-1, 1) \), which corresponds to increasing \( \Delta \) and decreasing \( \lambda \). In this case, \( \partial_{\mathbf{d}} D = \frac{1}{\sqrt{2}}(\partial_{\Delta} - \partial_{\lambda}) D \). Distinct phase regions are revealed in the color contour and are separated by discontinuities in \( \partial_{\mathbf{d}} D \) (white dashed lines).

We then show the phase diagrams in the (\( \lambda, \Delta \)) plane for different values of \( V \) in Fig.~\ref{fig:phase diagrams}(d).
For \( V=1.5 \) (leftmost panel), the entire parameter space is occupied by the FM phase (dark blue).
When \( V=3 \) (second panel to the left), the AFM phase emerges (cyan) at lower \( \lambda \) values. As \( V \) increases further (the rightmost two panels), a third region, the LC phase (red) appears at small \( \lambda \), particularly around \( \Delta \sim 2 \). With increasing \( V \), both phase boundaries shift toward larger \( \lambda \) values, demonstrating the effect of interaction on the dynamic phase transitions.
These are consistent with the understanding that small $V$ or large $\lambda$ tends to suppress the nonlinear effects.

\section{Steady-state phase transition}
We now analyze the steady-state transition between the AFM and FM phases.
A well-known characteristic of quantum phase transitions in closed systems is the divergence of the ground-state correlation length near the critical point~\cite{SachdevQPT}. For the steady-state transition in open quantum systems, the correlation length may not diverge, but the transition still leaves signals in the correlation function~\cite{Nagy2011, Ali2024}.
To demonstrate this, we define the correlation function as
\begin{equation}
C_{r} = \frac 1 N\sum_i|\langle \sigma_i^z \sigma_{i+r}^z\rangle - \langle \sigma_i^z \rangle\langle\sigma_{i+r}^z\rangle|,
\end{equation}
where \( \langle \cdot \rangle \) is the expectation value of steady state \(\rho_s\).

\begin{figure}[tbp]
\centering
\includegraphics[width=\columnwidth]{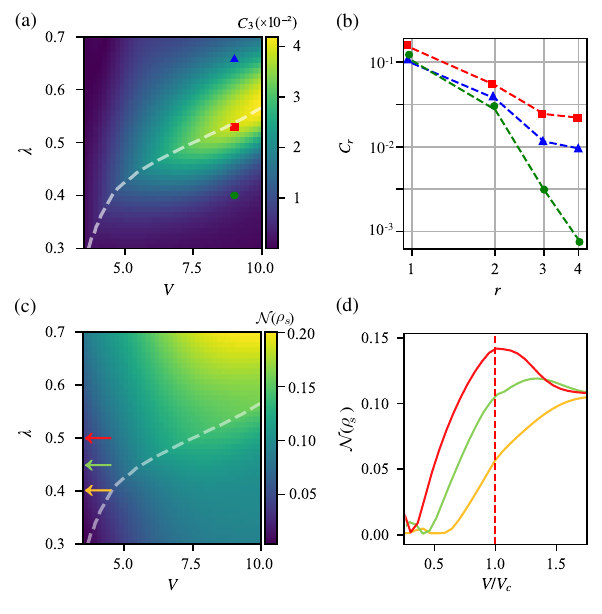}
\caption{Steady-state correlation functions and entanglement negativity.
(a) Variation of the steady-state correlation function  $C_{r=3}$ as a function of system parameters, where the color represents \( C_{3} \), and the dashed line denotes the AFM--FM phase boundary.
(b) Log-log plot of the correlation function \( C_{r} \) as a function of \( r \), where different colored curves correspond to markers of the corresponding color in (a).
(c) Steady-state entanglement negativity \( \mathcal{N}(\rho_s) \) as a function of system parameters, with the dashed line indicating the AFM--FM phase boundary.
(d) Variation of entanglement negativity with \( V/V_c \) for \( \lambda= 0.4, 0.45, 0.5 \), where \( V_c \) represents the AFM--FM transition point. Different curves correspond to distinct values of \( \lambda \), as indicated by the arrows of matching colors in (c).
We set \( N=6 \) in (a)(c)(d), and \( N=8 \) in (b).
The remaining parameters are fixed as \( \Omega=1.5 \), \( \Delta = 2 \), with periodic boundary conditions.
}
\label{fig:quantum pt}
\end{figure}

In Fig.~\ref{fig:quantum pt}(a), we numerically evaluate the steady-state correlation function \(C_{r=3}\) with periodic boundary condition for \( N=6 \).
Apparently, the correlation function peaks near the AFM-FM phase boundary (white dashed).
To further explore this behavior, we show a log-log plot of the correlation \( C_{r} \) as a function of $r$ for a system size of \( N=8 \), as shown in Fig.~\ref{fig:quantum pt}(b). The dashed line with red square markers corresponds to parameters at the phase boundary, while the other two curves represent parameters on either side of the phase boundary [see corresponding markers Fig.~\ref{fig:quantum pt}(a)].
Notably, the correlation function at the phase boundary exhibits an approximately linear behavior in the log-log plot with distance $r$, consistent with an algebraic decay typical of quantum criticality.
Furthermore, by increasing $\lambda$ across the phase boundary, the variation of $C_r$ for a given $ r$ is not monotonic, peaking near the transition point.
On the other hand, when the phase boundary is well within the classical regime with small \( \lambda \), the correlation function remains nearly zero and does not exhibit significant changes across the phase transition. This is expected, since for small \( \lambda \), the coherent Hamiltonian is dominated by its mean-field component, leading to the relation \( \langle \sigma_i^z \sigma_{i+r}^z\rangle \approx \langle \sigma_i^z \rangle\langle\sigma_{i+r}^z\rangle \).

In Fig.~\ref{fig:quantum pt}(c), we calculate the entanglement negativity~\cite{Vidal2002, YALee2013,Nielsen2010} of the steady state \( \rho_s \), which is defined as
\begin{equation}\label{eqNeg}
\mathcal{N}(\rho_s) = \sum_{\alpha =\text{odd},\text{even}}\left(\sum_{\varepsilon^\alpha_i<0}|\varepsilon^{\alpha}_i | \right),
\end{equation}
where \( \{\varepsilon_i^{\text{odd}}\} (\{\varepsilon_i^{\text{even}}\})\) are the eigenvalues of the partially transposed density matrix \(  \rho_s^{\text{T}_{\text{odd}}} (\rho_s^{\text{T}_{\text{even}}})\), where T\( _{\text{odd}} (\text T_{\text{even}}) \) denotes the partial transpose operation that transposes the degrees of freedom of all odd (even) lattice sites in the steady-state density matrix.
The summation over $\alpha$ in Eq.~(\ref{eqNeg}) accounts for the difference between odd and even sites in the AFM phase.
By fixing \( \lambda \) and varying the interaction strength \( V \), we observe that, for larger values of \( \lambda \), the entanglement negativity shows a peak at the phase transition \( V_c \), which is a key feature of the steady-state quantum phase transition~\cite{Nagy2011}. However, no such peak is observed for smaller \( \lambda \), so that the entanglement negativity is not an effective indicator for phase boundaries at small $\lambda$.
This is evident in Fig.~\ref{fig:quantum pt}(d), where the three curves correspond to \( \lambda = 0.4, 0.45,\) and \( 0.5 \), as indicated by the arrows of corresponding colors in Fig.~\ref{fig:quantum pt}(c). 
Such a behavior is consistent with that of the correlation function, indicating that the critical character of the steady-state phase transition is more pronounced as the system is tuned toward the quantum limit (with larger $\lambda$).

\section{Frequency Bifurcation}
For the dynamics near the classical limit, a better understanding can be obtained from the perspective of frequency bifurcation.
In nonlinear dynamics, frequency bifurcation is a common phenomenon, which is often associated with nonlinear interactions, external drives, and parameter modulations.
Typically, when a parameter exceeds a critical value, a Hopf bifurcation occurs, leading to new oscillatory patterns. Furthermore, period-doubling or quasiperiodic bifurcations can also appear, eventually leading to chaotic dynamics~\cite{Strogatz1994}.
In our system, similar frequency bifurcation emerge as the long-time dynamics of the system evolve from limit cycles to the AFM steady state.

Spectral analysis is one of the most common methods used to investigate oscillatory behavior in nonlinear systems~\cite{Strogatz1994}. In the previous sections, we observed that, for small values of $\lambda$, the long-time dynamics exhibit oscillatory behavior characteristic of limit cycles. Here we study the frequency spectrum \( \tilde n(f) \) of the spin-up occupation, obtained by performing a fast Fourier transform (FFT) on the time series of \( n_{\text{even}}^r(t) \). The FFT is applied to the time series taken within a time window following a sufficiently long evolution. This yields the frequency distribution \( \tilde n(f) \), where \( f \) denotes frequency.

\begin{figure}[htb]
\centering
\includegraphics[width=1.0\columnwidth]{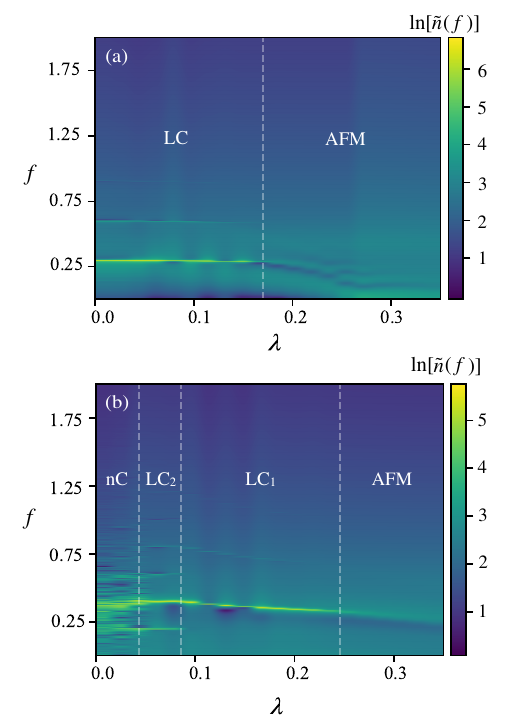}
\caption{
Frequency spectrum of the long-time density matrix, with \( \Delta = 2 \), \( \Omega = 1.5 \), and \( N = 6 \). (a) The case of \( V = 5 \), where only the LC and AFM phases are present. The vertical dashed line marks the Hopf bifurcation separating the two phases.
(b) The case of \( V = 9 \), where rich nonlinear behaviors emerge, as four distinct phases can be identitifed: the near-chaotic phase (nC), the period-doubled limit cycle phase (LC\(_{2}\)), the single-period limit cycle phase (LC\(_{1}\)), and the AFM phase. The dashed lines indicate the boundaries between these phases. Only the positive-frequency components are shown, and the color represents the logarithmic amplitude \(\ln[\tilde n(f)]\).}
\label{fig:frequency}
\end{figure}

In Fig.~\ref{fig:phase diagrams}(c)(d), the transition between the LC and AFM phases is characterized by discontinuities in the derivative of $D$. An alternative and more standard identifier is through spectral analysis.
In Fig.~\ref{fig:frequency}(a), we compute the logarithmic amplitude \( \ln[\tilde n(f)] \) as a function of frequency \( f \) (only the positive-frequency components are shown) and the interpolation parameter \( \lambda \), with \( V = 5 \) and \( \Delta = 2 \). We observe that for \( \lambda \lesssim 0.17 \) (vertical dashed line), there exists a dominant frequency component, a prominent signature of oscillations in the LC phase. For \( \lambda > 0.17 \), this frequency component disappears, signifying the system approaching a steady state at long times.
This transition corresponds to a Hopf bifurcation, where the limit-cycle oscillations vanish and the system settles into a stable fixed point~\cite{Strogatz1994}.

As we increase $V$ further, the frequency spectrum begins to show more complex nonlinear behaviors. In Fig.~\ref{fig:frequency}(b), we show \( \ln[\tilde n(f)] \) with a larger interaction strength \( V = 9 \).
For small \( \lambda \), in the region labeled ``nC'', the spectrum shows a wide, almost continuous range, typical of near-chaotic motion. This spectral broadening indicates that the system no longer exhibits well-defined periodic dynamics.
As \( \lambda \) increases, the distribution of the frequency components becomes sparser. Around \( \lambda=0.04 \), the spectrum is composed of the main frequency, and a frequency at half of the main one, indicating that the system is entering the LC phase.
We label this region as ``LC$_{2}$''.
As \( \lambda \) increases and crosses the vertical dashed line near \( \lambda \sim 0.08 \), a period-doubling bifurcation occurs, the half-frequency components disappear, and the oscillation period is reduced by half. This marks a transition into the ``LC$_1$'' region, where the system shows single-frequency limit-cycle dynamics.
Finally, at \( \lambda \sim 0.25 \), a Hopf bifurcation occurs, where oscillatory dynamics disappear and the system enters the AFM state. Here what remains from the LC phase is only the spontaneously broken translational symmetry [see also Fig.~\ref{fig:example}(b)(d)]

\section{Summary}
In summary, we propose an ensemble averaged cluster model with an interpolated coherent Hamiltonian to study the long-time dynamics of a dissipative Ising chain. Starting from the classical limit where the long-time dynamics exhibit either limit cycles or near chaotic behaviors, we show how the system sequentially undergoes frequency bifurcations and a steady-state phase transition, so that the persistent oscillations of the limit cycles change into steady-state approaching dynamics. Our model is relevant to dissipative Rydberg gases, wherein the interpolation parameter $\lambda$, which qualitatively measures the quantum correlations within the system.
For future studies, it would be desirable to apply a similar approach to dissipative Ising models in higher dimensions.

\begin{acknowledgments}
This work is supported by the National Natural Science Foundation of China (Grant No. 12374479), and the Innovation Program for Quantum Science and Technology (Grant No. 2021ZD0301205).
\end{acknowledgments}

\section*{Data Availability}
The data that support the findings of this study are openly available~\cite{zhang2025code}.

\end{document}